\documentclass[conference]{IEEEtran}
\IEEEoverridecommandlockouts

\usepackage{amsmath, mathrsfs, graphicx,epsfig,amssymb, amsfonts, color, subfigure}
\usepackage[center]{caption}

\newtheorem{lemma}{\bf Lemma}

\newtheorem{defn}{\bf Definition}
\newtheorem{thm}{\bf Theorem}

\newtheorem{rem}{\bf Remark}
\newtheorem{cor}{\bf Corollary}
\newtheorem{ex}{\bf Example}

\begin{document}

\title{\huge{The Generalized Degrees of Freedom of the MIMO Interference Channel}}

\author{{Sanjay Karmakar ~~~~ Mahesh K.~Varanasi}
\thanks{This work was supported in part by NSF Grant CCF-0728955.
The authors are with the Department of Electrical, Computer, and
Energy Engineering, University of Colorado, Boulder, CO 80309-0425
USA (e-mail: {sanjay.karmakar, varanasi}@colorado.edu).} }

\maketitle

\begin{abstract}
The generalized degrees of freedom (GDoF) region of the MIMO Gaussian interference channel is obtained for the general case with an arbitrary number of antennas at each node and where the SNR and interference-to-noise ratios (INRs) vary with arbitrary exponents to a nominal SNR. The GDoF region reveals various insights through the joint dependence of optimal interference management techniques at high SNR on the SNR exponents that determine the relative strengths of direct-link SNRs and cross-link INRs and the numbers of antennas at the four terminals. For instance, it permits an in-depth look at the issue of rate-splitting and partial decoding at high SNR and it reveals that, unlike in the SISO case, treating interference as noise is not GDoF optimal always even in the very weak interference regime. Moreover, while the DoF-optimal strategy that relies just on transmit/receive zero-forcing beamforming and time-sharing is not GDoF optimal (and thus has an unbounded gap to capacity) the precise characterization of the very strong interference regime, where single-user DoF performance can be achieved simultaneously for both users, depends on the relative numbers of antennas at the four terminals and thus deviates from what it is in the SISO case. For asymmetric numbers of antennas at the four nodes the shape of the symmetric GDoF curve can be a ``distorted W" curve to the extent that for certain MIMO ICs it is a ``V" curve.
\end{abstract}
\IEEEpeerreviewmaketitle

\section{Introduction}
The GDoF region metric was introduced in \cite{ETW1}. It generalizes the usual notion of the conventional degrees of freedom (DoF) region metric by additionally emphasizing the signal level as a signaling dimension. It therefore characterizes the simultaneously accessible fractions of spatial and signal-level dimensions (per channel use) by the two users in the limit of high signal-to-noise ratio (SNR) while the ratios of the SNRs and INRs (interference-to-noise ratios) relative to a reference SNR, each expressed in the {\em dB scale}, are held constant, with each constant taken, in the most general case, to be arbitrary. The GDoF region was obtained for the SISO IC in \cite{ETW1} and the symmetric GDoF (the maximum common GDoF achievable by each of the two users) for the symmetric SISO IC (with equal SNRs and equal INRs for the two users, i.e, with ${\rm INR} = {\rm SNR}^\alpha $) was evaluated to be the well-known ``W" curve which clearly delineates the very weak, weak, moderate, strong and very strong interference regimes, depending on the value of $\alpha$, thereby shedding light on interference management techniques as a function of the severity or mildness of the interference.

There have been several other recent works on characterizing the GDoF of various channels. For example, in \cite{PBT}, the symmetric GDoF of a class of symmetric MIMO ICs -- for which the SNRs at each receiver are the same and the INRs at each receiver are also the same, with ${\rm INR} = {\rm SNR}^\alpha $-- and where both transmitters have $M$ antennas and both receivers have $N$ antennas, with
the restriction $N \geq M$, was obtained and found to be a ``W" curve also. In \cite{Jafar_Vishwanath_GDOF}, the symmetric GDoF in the perfectly symmetric (with all direct links having identical gains and all cross links having identical gains) scalar $K$-user interference network was found (see also \cite{Bandemer-Cyclic}). In \cite{Gou_Jafar}, the symmetric GDoF was obtained for the $(N+1)$-user symmetric SIMO IC with $N$ antennas at each receiver and with equal direct link SNRs and equal cross link INRs. The symmetric GDoF of a symmetric model of the scalar X-channel was found in \cite{Huang_Cadambe_Jafar}.

The GDoF result of this paper generalizes that in \cite{ETW1} to the MIMO IC with an arbitrary number of antennas at each node. It recovers the symmetric GDoF result of \cite{PBT} for the class of symmetric MIMO ICs considered therein. The achievability scheme considered here, unlike that in \cite{PBT}, is GDoF optimal in the most general case. It also generalizes the DoF region result obtained in \cite{JFak} for the MIMO IC with an arbitrary number of antennas at each node. The GDoF result of this paper provides several insights that include whatever is common between symmetric MIMO ICs and SISO ICs and what is not, but more importantly, it gives rise to new insights into optimal signaling strategies that make jointly optimal use of the available spatial and signal level dimensions.

The rest of the paper is organized as follows. In Section~\ref{sec_channel_model_and_preliminaries} we describe the channel model and the GDoF optimal coding scheme. The GDoF region of the general MIMO IC will be stated as the main result of the paper in Section~\ref{sec_main_result} along with its invariability with respect to direction of information flow. In Section~\ref{sec_insights} the various insights revealed by the GDoF analysis are given. Section~\ref{sec_conclusion} concludes the paper.

\section{Channel Model and preliminaries }
\label{sec_channel_model_and_preliminaries}
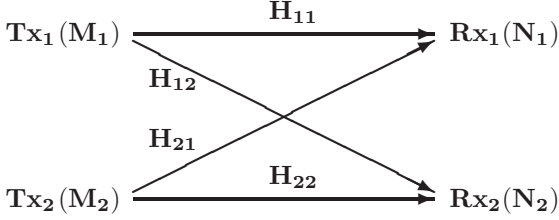
\begin{figure}[!thb]
\setlength{\unitlength}{1mm}
\begin{picture}(80,40)
\thicklines
\put(20,10){\vector(1,0){40}}
\put(20,32){\vector(1,0){40}}
\put(20,11){\vector(2,1){40}}
\put(20,31){\vector(2,-1){40}}
\put(10,9){$\mathbf{(M_2)}$}
\put(69,9){$\mathbf{(N_2)}$}
\put(10,31){$\mathbf{(M_1)}$}
\put(69,31){$\mathbf{(N_1)}$}

\put(03,9){$\mathbf{Tx_2}$}
\put(62,9){$\mathbf{Rx_2}$}
\put(03,31){$\mathbf{Tx_1}$}
\put(62,31){$\mathbf{Rx_1}$}

\put(38,12){$\mathbf{H_{22}}$}
\put(38,34){$\mathbf{H_{11}}$}
\put(22,17){$\mathbf{H_{21}}$}
\put(22,25){$\mathbf{H_{12}}$}
\end{picture}
\caption{The $(M_1,N_1,M_2,N_2)$ MIMO IC.}
\label{channel_model_two_user_IC}
\end{figure}

The 2-user MIMO IC, with $M_i$ and $N_i$ antennas at transmitter $i$ ($Tx_i$) and receiver $i$ ($Rx_i$), respectively, for $i=1,~2$ as shown in figure~\ref{channel_model_two_user_IC} (hereafter referred to as $(M_1,N_1,M_2,N_2)$ IC) is considered. We shall consider a time-invariant or fixed channel where the channel matrices, $H_{ij}$'s remain fixed for the entire duration of communication and whose entries are drawn i.i.d. from a continuous distribution, which ensures that they are full rank with probability one (w.p.1). We also incorporate a real-valued attenuation factor, denoted as $\eta_{ij}$, for the signal transmitted from $Tx_i$ to receiver $Rx_j$. At time $t$, $Tx_i$ chooses a vector ${X}_{it}\in \mathbb{C}^{M_i\times 1}$ and sends $\sqrt{P_i}{X}_{it}$ over the channel, where we assume the following average input power constraint at $Tx_i$,
\begin{equation}
\label{power_constraint}
\frac{1}{n}\sum_{t=1}^{n}\textrm{Tr} (Q_{it}) ~\leq ~1,
\end{equation}
for $ i \in \{ 1,2 \}$, where $ Q_{it}=\mathbb{E}({X}_{it}{X}_{it}^{\dagger})$.
Note that in the above power constraint $Q_{it}$'s can depend on the channel matrices.

The received signals at time $t$ can be written as
\begin{eqnarray}
\label{system_eq_two_user_IC2}
Y_{1t}=\sqrt{\rho_{11}}H_{11}X_{1t}+ \sqrt{\rho_{21}}H_{21}X_{2t}+Z_{1t},\\
Y_{2t}= \sqrt{\rho_{22}}H_{22}{X}_{2t}+\sqrt{\rho_{12}} H_{12}X_{1t}+Z_{2t},
\end{eqnarray}
where $Z_{it}\in\mathbb{C}^{N_i\times 1}$ are i.i.d $\mathcal{CN}(\mathbf{0}, I_{N_i})$ across $i$ and $t$, $\rho_{ii}=\eta_{ii}\sqrt{P_i}=\rho^{\alpha_{ii}}$ represents the signal-to-noise ratio (SNR) at receiver $i$ and $\rho_{ij}=\eta_{ij}\sqrt{P_i}=\rho^{\alpha_{ij}}$ represents the interference-to-noise ratio (INR) at receiver $j$ for $i \neq j\in \{1,2\}$. In what follows, the MIMO IC with channel matrices, SNRs and INRs as described above will be denoted by $\mathcal{IC}\left(\mathcal{H},\bar{\rho}\right)$, where $\mathcal{H}=\{H_{11},H_{12},H_{21},H_{22}\}$ and $\bar{\rho}=[\rho_{11},\rho_{12},\rho_{21},\rho_{22}]$ or equivalently as $\mathcal{IC}\left(\mathcal{H},\bar{\alpha}\right)$ where $\bar{\alpha}=[\alpha_{11},\alpha_{12},\alpha_{21},\alpha_{22}]$. The capacity region of $\mathcal{IC}\left(\mathcal{H},\bar{\alpha}\right)$ is defined in the usual way (cf. \cite{Sanjay_Varanasi_Cap_MIMO_IC_const_gap}) and will be denoted by $\mathcal{C}\left(\mathcal{H},\bar{\alpha}\right)$. The GDoF region is defined next.

\begin{defn}
The GDoF region, $\mathcal{D}_o(\bar{M},\bar{\alpha})$, of $\mathcal{IC}(\mathcal{H},\bar{\alpha})$ is defined as
\begin{IEEEeqnarray}{rl}
\mathcal{D}_o(\bar{M}, \bar{\alpha})=\Big\{(d_1,d_2): &~d_i=\lim_{\rho_{ii}\to \infty}\frac{R_i}{\log(\rho_{ii})}, i\in\{1,2\}\nonumber \\
\label{def_GDOF}
&~\textrm{such that}~(R_1,R_2)\in \mathcal{C}(\mathcal{H},\bar{\alpha}) \Big\}.
\end{IEEEeqnarray}
\end{defn}

To derive the GDoF region we shall use a recent constant-gap-to-capacity result found by the authors in \cite{Sanjay_Varanasi_Cap_MIMO_IC_const_gap}. Before stating the result, we describe the coding scheme whose rate region is within a constant gap to the capacity region and which therefore also achieves the fundamental GDoF region of the channel.
\begin{defn}[A simple HK coding scheme]
\label{def_coding_scheme}
Each user divides its message into two sub-messages (called the private and public messages hereafter) and uses superposition coding to
encodes the two sub-messages with mutually independent random zero-mean Gaussian code books so that we have
\begin{equation}
\begin{array}{c}
X_1={U}_1+{W}_1,\\
X_2={U}_2+{W}_2,
\end{array}
\end{equation}
where ${U}_i$ and ${W}_i$ represent the codewords of the private and public messages of user $i$, respectively. Moreover, the covariance matrices of the public and private messages are taken for each $i \in \{1,2\}$ to be
\begin{IEEEeqnarray}{rl}
\label{eq_power_split}
K_{iu}(H)\triangleq\mathbb{E}(U_i U_i^{\dagger}) =& \frac{1 }{M_{i}}\left(I_{M_i}+\rho_{ij} H_{ij}^{\dagger}H_{ij}\right)^{-1},\nonumber \\
K_{iw}(H)\triangleq\mathbb{E}(W_i W_i^{\dagger})=&  \left(\frac{I_{M_i}}{M_{i}}-K_{iu}\right).
\end{IEEEeqnarray}
In what follows, we shall refer to such a coding scheme as the $\mathcal{HK}\left(\{K_{1u}(H),K_{1w}(H),K_{2u}(H),K_{2w}(H)\}\right)$ scheme. Using the singular value decomposition of $H_{ij}=V_{ij}\Sigma_{ij}U_{ij}^\dagger$, where $V_{ij}$ and $U_{ij}$'s are unitary matrices and $\Sigma_{ij}$ is a rectangular matrix containing the singular values, it can be shown that $X_i$ can be written as
\begin{IEEEeqnarray}{rl}
\label{eq_structure_of_streams}
X_i=\sum_{k=1}^{m_{ij}}\sqrt{1-\frac{1}{G_i}}&x_{ic}^{(k)}U_{ij}^{[k]}+\sum_{l=1}^{m_{ij}}\sqrt{\frac{1}{G_i}}x_{ip}^{(l)}U_{ij}^{[l]}+\nonumber \\&\sum_{m=1+m_{ij}}^{M_i}\frac{1}{\sqrt{M_i }}x_{ip}^{(m)}U_{ij}^{[m]},
\end{IEEEeqnarray}
where $G_i=M_i(1+\rho_{ij} \lambda_{ij}^{(k)})$, $m_{ij}=\min\{M_i,N_j\}$ and the quantities $U_{ij}^{[k]}$, $\lambda_{ij}^{(k)}$, $x_{ic}^{(k)}$ and $x_{ip}^{(k)}$ represent the $k^{th}$ column of $U_{ij}$, the $k^{th}$ eigenvalue of $H_{ij}^\dagger H_{ij}$ and the symbols of the $k^{th}$ stream of the public and private messages, respectively. For all $1\leq i\leq 2,1\leq k\neq l\leq M_i$, we have
\begin{IEEEeqnarray*}{rl}
\mathbb{E}\left[{|x_{ic}^{(k)}|}^2\right]=1,~\mathbb{E}\left[{|x_{ip}^{(k)}|}^2\right]=1, ~\mathbb{E}\left[\left(x_{ic}^{(k)}(x_{ip}^{(l)})^*\right)\right]=0.
\end{IEEEeqnarray*}
\end{defn}

\begin{rem}
Each $U_{ij}^{[k]}$ is a right singular vector corresponding to a zero singular value of the matrix $H_{ij}$. Hence note that none of the $x_{ip}^{(k)}$'s for $(m_{ij}+1)\leq k\leq M_i$ reach the useful signal space of $Rx_j$. On the other hand, each of $x_{ip}^{(l)}$ for $1\leq l\leq m_{ij}$ is transmitted at a power of $\rho^{-\alpha_{ij}}$ and hence reaches $Rx_j$ at the noise floor. This explains why all of $U_i$ is called the private message of user $i$.
\end{rem}

It was shown by the authors in \cite{Sanjay_Varanasi_Cap_MIMO_IC_const_gap} that this coding scheme can achieve a rate region, $\mathcal{R}_a(\mathcal{H},\bar{\rho})$ on the MIMO IC which is within a constant number of bits to an outer bound $\mathcal{R}^u(\mathcal{H},\bar{\rho})$ of the capacity region of the channel. An explicit expression for both $\mathcal{R}^u(\mathcal{H},\bar{\rho})$ and $\mathcal{R}_a(\mathcal{H},\bar{\rho})$ can be found in Lemma~1 and 3 of \cite{Sanjay_Varanasi_Cap_MIMO_IC_const_gap}, respectively, and are not provided here. Since a constant number of bits is insignificant in the GDoF analysis, the $\mathcal{C}(\mathcal{H},\bar{\alpha})$ in the definition of the GDoF region can be replaced by either $\mathcal{R}^u(\mathcal{H},\bar{\rho})$ or $\mathcal{R}_a(\mathcal{H},\bar{\rho})$ to compute the GDoF region of the MIMO IC. We state this fact as a lemma for easy further reference.
\begin{lemma}
\label{lem_alternate_def_GDOF}
The GDoF region of the MIMO IC is given as
\begin{IEEEeqnarray}{rl}
\mathcal{D}_o(\bar{M},\bar{\alpha})=\Big\{&(d_1,d_2): ~d_i=\lim_{\rho_{ii}\to \infty}\frac{R_i}{\log(\rho_{ii})}, i\in\{1,2\}\nonumber \\
\label{eq_alternate_def_GDOF}
&~~\textrm{and}~(R_1,R_2)\in \mathcal{R}^u(\mathcal{H},\bar{\alpha})  \Big\},
\end{IEEEeqnarray}
where $\mathcal{R}^u(\mathcal{H},\bar{\alpha})=\mathcal{R}^u(\mathcal{H},\bar{\rho})$ is given by Lemma~1 of \cite{Sanjay_Varanasi_Cap_MIMO_IC_const_gap}.
\end{lemma}

To describe the GDoF region of the MIMO IC we need the following definitions.
\begin{defn}
For any $u\in \mathbb{R}$ and $(a_i,u_i)\in \mathbb{R}^{+2}$ for $i\in \{1,2\}$, $f\left(u,(a_1,u_1),(a_2,u_2)\right)=$
\begin{IEEEeqnarray}{l}
\label{eq_def_f}
\left\{\begin{array}{c}
\min\{u,u_1\}a_1+\min\{(u-u_1)^+,u_2\}a_2,~\textrm{if}~a_1\geq a_2;\\
\min\{u,u_2\}a_2+\min\{(u-u_2)^+,u_1\}a_1,~\textrm{if}~a_1< a_2.
\end{array}\right.
\end{IEEEeqnarray}
\end{defn}

\begin{defn}
For any $u\in \mathbb{R}$ and $(a_i,u_i)\in \mathbb{R}^{+2}$ for $i\in \{1,2,3\}$,
$g\left(u,(a_1,u_1),(a_2,u_2),(a_3,u_3)\right)=$
\begin{IEEEeqnarray}{rl}
\min\{u,u_{i_1}\}a_{i_1}+&\min\{(u-u_{i_1})^+,u_{i_2}\}a_{i_2}+\nonumber\\
\label{eq_def_g}
&\min\{(u-u_{i_1}-u_{i_2})^+,u_{i_3}\}a_{i_3},
\end{IEEEeqnarray}
for $i_1,i_2,i_3\in \{1,2,3\}$ such that $a_{i_1}\geq a_{i_2}\geq a_{i_3}$.
\end{defn}

\begin{rem}
\label{rem_MAC_interpretation}
The function g(.) in the above definition can be interpreted as the sum DoFs achievable on a 3-user MIMO multiple-access channel (MAC) with $u$ antennas at the receiver, $u_i$ antennas at the $i^{th}$ transmitter, where the SNR of the $i^{th}$ user is $\rho^{a_i}$ for $i\in \{1,2,3\}$. 
Similarly, $f(.)$ can be interpreted as the sum GDoF achievable on a 2-user MIMO MAC.
\end{rem}

\section{The GDoF region of the MIMO IC}
\label{sec_main_result}
Using the explicit expression for the upper bounds to the capacity region of the MIMO IC, $\mathcal{R}^u(\mathcal{H},\bar{\alpha})$ from Lemma~1 of \cite{Sanjay_Varanasi_Cap_MIMO_IC_const_gap} and using it in Lemma~\ref{lem_alternate_def_GDOF} we get the main result of this paper.
\begin{thm}
\label{thm_mainresult}
The GDoF region of $\mathcal{IC}(\bar{M},\bar{\alpha})$ is the set of DoF tuples $(d_1,d_2)$, denoted by $\mathcal{D}_o(\bar{M},\bar{\alpha})$, where $d_i\in \mathbb{R}^{+}$ for $i=1,2$ satisfy the following conditions:
\begin{IEEEeqnarray*}{rl}
\label{eq_gdofr1}
d_1\leq &\min\{M_1, N_1\};\\
d_2\leq &\min\{M_2, N_2\};\\
\label{eq_gdofr3}
d_3\leq & f\left(N_2, (\alpha_{12},M_1),(\alpha_{22},M_2)\right)+\\
& f\left(N_1, (\beta_{12},m_{12}),(\alpha_{11},(M_1-N_2)^+)\right);\\
\label{eq_gdofr4}
d_4\leq & f\left(N_1, (\alpha_{21},M_2),(\alpha_{11},M_1)\right)+\\
&f\left(N_2, (\beta_{21},m_{21}),(\alpha_{22},(M_2-N_1)^+)\right);\\
d_5\leq & g\left(N_1,(\alpha_{21},M_2),(\beta_{12},m_{12}),(1,(M_1-N_2)^+)\right)+\nonumber \\
\label{eq_gdofr5}
&g\left(N_2,(\alpha_{12},M_1),(\beta_{21},m_{21}),(\alpha_{22},(M_2-N_1)^+)\right);\\
d_6\leq & f\left(N_1, (\alpha_{21},M_2),(\alpha_{11},M_1)\right)+\\
&f\left(N_1, (\beta_{12},m_{12}),(\alpha_{11},(M_1-N_2)^+)\right)+\nonumber\\
\label{eq_gdofr6}
&g\left(N_2,(\alpha_{12},M_1),(\beta_{21},m_{21}),(\alpha_{22},(M_2-N_1)^+)\right);\\
d_7\leq & f\left(M_2, (\alpha_{21},N_1),(\alpha_{22},N_2)\right)+\\
&f\left(N_2, (\beta_{21},m_{21}),(\alpha_{22},(M_2-N_1)^+)\right)+\nonumber\\
\label{eq_gdofr7}
&g\left(N_1,(\alpha_{21},M_2),(\beta_{12},m_{12}),(1,(M_1-N_2)^+)\right),
\end{IEEEeqnarray*}
where $d_{3,4,5}=d_1+\alpha_{22}d_2$, $d_6=(2d_1+\alpha_{22}d_2)$, $d_7=(d_1+2\alpha_{22}d_2)$, $\beta_{ij}=(\alpha_{ii}-\alpha_{ij})^+$, functions $f(.,.,.)$ and g(.,.,.,.) are as defined in equation \eqref{eq_def_f} and \eqref{eq_def_g}, respectively, and $m_{ij} \triangleq \min\{M_i,N_j\}$ for $i\neq j\in \{1,2\}$.
\end{thm}

\begin{rem}[Interpretation of the different bounds]
We know that the GDoF optimal coding scheme divides each user's message into two sub-messages. Let the DoFs of the private and the public messages of user $i$ be denoted by $d_{ip}$ and $d_{ic}$, respectively. Clearly, $Tx_1$ can send $(M_1-N_2)^+\alpha_{11}$ private DoFs to its desired user through the null space of the channel $H_{12}$ and a maximum of $m_{12}(\alpha_{11}-\alpha_{12})^+$ DoFs at a power level of $\rho^{-\alpha_{12}}$ so that it reaches $Rx_2$ below noise floor. This implies that $d_{1p}$ is upper bounded as follows:
\begin{equation*}
d_{1p}\leq f\left(N_1, ((\alpha_{11}-\alpha_{12})^+,m_{12}),(\alpha_{11},(M_1-N_2)^+)\right).
\end{equation*}
On the other hand, since $d_{1c}$ is decoded at $Rx_2$, $Rx_2$ is a MAC receiver with respect to $W_1$ and $X_2$ and thus we have (recall Remark~\ref{rem_MAC_interpretation})
\begin{equation*}
    (d_{1c}+\alpha_{22} d_2)\leq f\left(N_2, (\alpha_{12},M_1),(\alpha_{22},M_2)\right).
\end{equation*}
Combining the above two equations we get the $3^{rd}$ bound of the GDoF region. All the other bounds of the region can be similarly explained.
\end{rem}

Note that the GDoF region is defined through seven different inequalities and hence in general for any given values of $M_i$, $N_j$ and $\bar{\alpha}$, the GDOF region can be a nine-sided polygon (including the $d_1$ and $d_2$ axis).

\begin{figure}[htp]
 \begin{center}
\includegraphics[scale=.3]{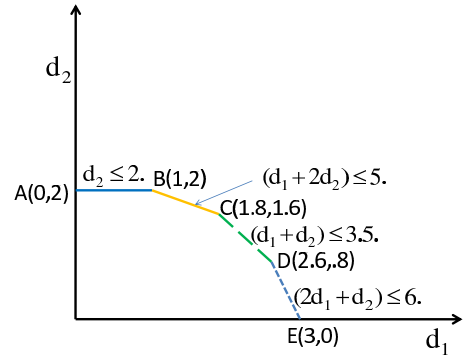}
\end{center}
\caption{GDoF region of the $(3,3,2,2)$ IC. }
\label{fig_explicit_scheme}
\end{figure}

We know from the original HK coding scheme~\cite{Han_Kobayashi} that if a rate tuple $(R_1, R_2)$ lies in the HK achievable rate region, then there exists at least one choice for the rates of the private and public messages of each user such that they add up to $R_i$. The same is true for DoF tuples, i.e., given a DoF tuple $(d_1,d_2) \in \mathcal{D}_o(\bar{M},\bar{\alpha})$, there exist at least one 4-tuple $(d_{1c}, d_{1p},d_{2c}, d_{2p})$ such that $d_i=(d_{ic}+d_{ip})$ is achievable for $i=1,2$. There is a general technique to choose such a DoF split, whose details are skipped here due to space constraints.

\begin{ex}
\label{ex_gdof_region_asymmetric}
Figure~\ref{fig_explicit_scheme} depicts the GDoF regions of the $(3,3,2,2)$ MIMO IC with $\bar{\alpha}=[1,\frac{2}{3},\frac{2}{3},1]$. In what follows, we shall explain the achievability of point B in Fig. \ref{fig_explicit_scheme} by $\mathcal{HK}(\{K_{1u},K_{1w},K_{2u},K_{2w}\})$. The distribution and power level of each user's private and public messages are specified in Definition~\ref{def_coding_scheme}; it only remains to specify the DoFs carried by the private and public messages, which are denoted by $d_{ip}$ and $d_{ic}$, respectively for a given $(d_1,d_2)$.
Using the general technique mentioned above for choosing the DoF split, it can be shown that for $(d_1,d_2)=(1,2)$, $(d_{1c}, d_{1p},d_{2c}, d_{2p}) = (0,1,1.2,.8)$ represents an achievable DoF quadruple in this example.

Since the first user needs to send only private information having DoF 1, it is best to send it in the direction of the null space of $H_{12}$, i.e.,
\begin{IEEEeqnarray}{l}
X_1=\frac{1}{\sqrt{3}}x_{1p}^{(3)}U_{12}^{[3]}.
\end{IEEEeqnarray}
On the other hand, the structure of the codeword for the second user is also clear from equation \eqref{eq_structure_of_streams}, i.e., $X_2$ is given as
\begin{IEEEeqnarray*}{rl}
\sum_{k=1}^{2}\frac{\sqrt{\rho_{21} \lambda_{21}^{(k)}}}{\sqrt{2 (1+\rho_{21} \lambda_{21}^{(k)})}}x_{2c}^{(k)}U_{21}^{[k]}+\sum_{l=1}^{2}\frac{1}{\sqrt{2 (1+\rho_{21} \lambda_{21}^{(l)})}}x_{2p}^{(l)}U_{21}^{[l]},
\end{IEEEeqnarray*}
where $x_{2c}^{(k)}$ and $x_{2p}^{(k)}$ carries $.6$ and $.4$ DoFs, respectively for both $k=1,2$.

{\it Decoding}: $Rx_1$ first projects the received signal on the 2 dimensional space which is perpendicular to $H_{11}U_{12}^{[3]}$ to remove the effect of $x_{1p}^{(3)}$ by zero forcing. In the resulting 2 dimensional signal space, only contribution from $W_2$ is present, which carries a DoF of $1.2$ and can be decoded because, the link from $Tx_2$ to $Rx_1$ is a $2\times 2$ point-to-point MIMO channel with effective SNR of $\rho^{.6}$. Once it is decoded $Rx_1$ removes its effect from the original received signal (the received signal before zero-forcing) and then it gets a clean channel from $Tx_1$ to itself and so it can decode $U_1$ as well. On the other hand, since $Rx_2$ does not face any interference\footnote{The interference that reaches below the noise floor is irrelevant in the GDoF computation.} from $Tx_1$, it can decode $W_2$ treating $U_2$ as noise. This is possible because treating $U_2$ as noise only raises the noise floor to $\rho^{.4}$ but the received signal power of $W_2$ is at $\rho$. Hence $Rx_2$ can decode $.6$ DoFs from each receive dimension. Next, subtracting the contribution of $W_2$ from the received signal, $Rx_2$ can decode $U_2$. The achievablity of any other point can be similarly explained. 
\end{ex}

\begin{cor}[Reciprocity of the GDoF region]
The GDoF region of the MIMO IC is same as that of its {\it reciprocal} channel i.e.,
\begin{equation*}\label{cor_reciprocity_GDOF}
\mathcal{D}_o(\bar{M},\bar{\alpha})=\mathcal{D}_o(\bar{M}^r,\bar{\alpha}^r),
\end{equation*}
where $\bar{M}^r=(N_1,M_1,N_2,M_2)$ and $\bar{\alpha}^r=[1,\alpha_{21},\alpha_{12},\alpha_{22}]$.
\end{cor}

In other words, the GDoF region of the channel does not change if the roles of the transmitters and the receivers are interchanged. This is a more general result than the reciprocity of the DoF region proved in \cite{JFak}.

\begin{rem}
The GDoF region result of \cite{ETW1} can be recovered by putting $M_1=M_2=N_1=N_2=1$.
The DoF region of the MIMO IC obtained in \cite{JFak} can be recovered by letting $\bar{\alpha}=[1,1,1,1]$ in Theorem~\ref{thm_mainresult}.
\end{rem}

\begin{rem}
The GDoF of the $(M,N,M,N)$ IC with $M\leq N$ and $\bar{\alpha}=[1,\alpha,\alpha,1]$ which was found in \cite{PBT} can be recovered by putting  $N_1=N_2=N, M_1=M_2=M$ and $\bar{\alpha}=[1,\alpha,\alpha,1]$, in Theorem~\ref{thm_mainresult}. The result of \cite{PBT} is valid only for $M\leq N$ and does not extend to the $M>N$ case. However, specializing Theorem~\ref{thm_mainresult} for $N_1=N_2=N, M_1=M_2=M>N$ and $\bar{\alpha}=[1,\alpha,\alpha,1]$ we get the following.
\end{rem}

\begin{figure}[htp]
  \begin{center}
    \subfigure[The fundamental GDoF region.]{\label{figure_gdof_MgeqN_channel}\includegraphics[scale=.28]{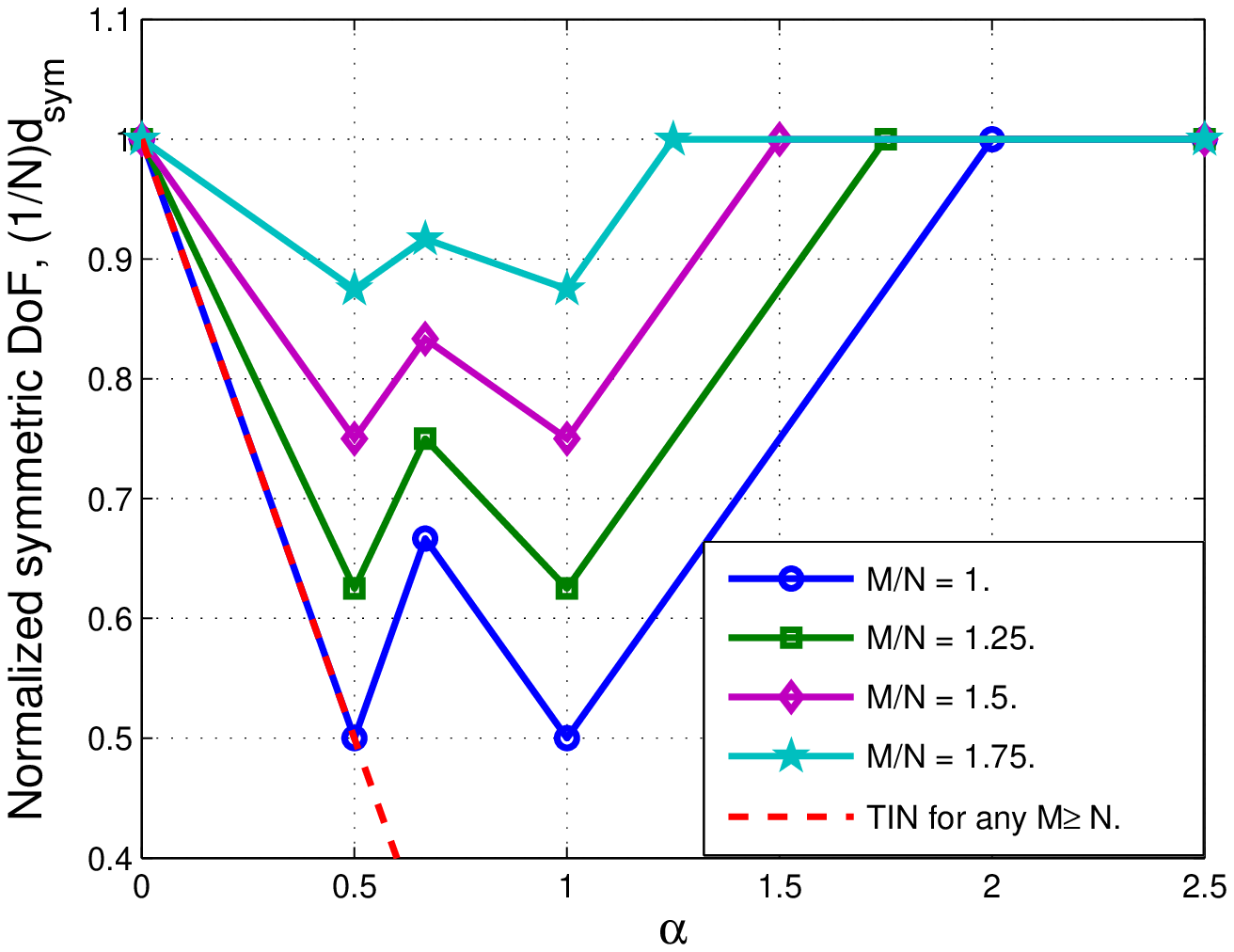}}
    \subfigure[GDoF region of the PBT scheme]{\label{figure_gdof_PBT_scheme}\includegraphics[scale=0.28]{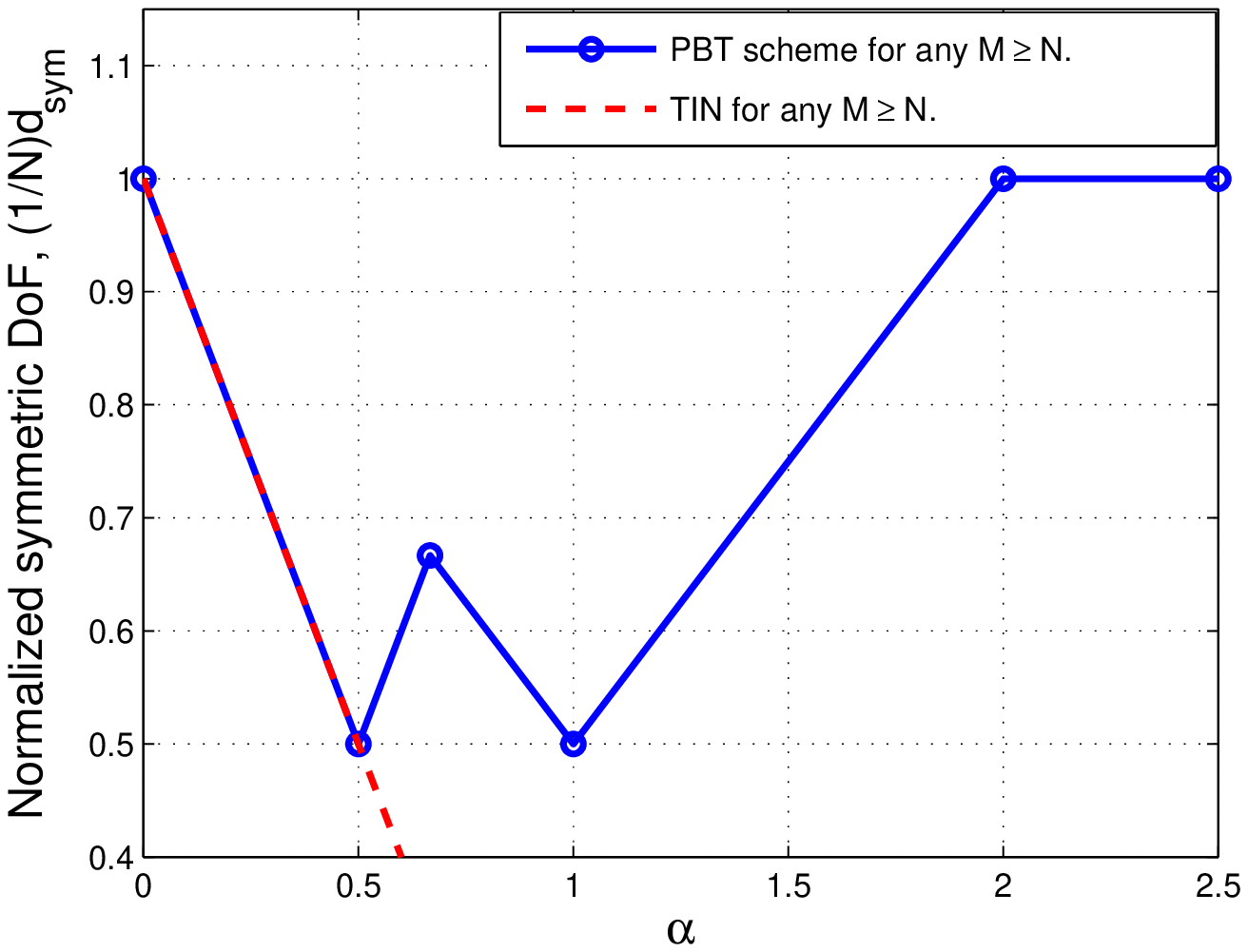}}
  \end{center}
\caption{Symmetric GDoF region of the $(M,N,M,N)$ IC. }
\end{figure}

\begin{cor}
\label{cor_gdof_symmetric_MgeqN}
The symmetric DoF ($d_{sym}=d_1=d_2$) of a $(M,N,M,N)$ IC with $M>N$ and $\bar{\alpha}=[1,\alpha,\alpha,1]$ is given by
\begin{equation*}
d_s(M,N,\alpha) \triangleq d_{sym}\leq \min \{N, D(\alpha)\}
\end{equation*}
where $D(\alpha)$ is given as
\begin{IEEEeqnarray}{l}
\label{eq_cor_MgN_case}
D(\alpha)=\left\{\begin{array}{ll}
N-(2N-M)\alpha, &0\leq \alpha<\frac{1}{2};\\
(M-N)+(2N-M)\alpha, &\frac{1}{2}\leq\alpha\leq \frac{2}{3};\\
N-\frac{\alpha}{2}(2N-M),  &\frac{2}{3}\leq\alpha\leq 1;\\
\frac{M}{2}+\frac{N}{2}(\alpha-1), &1\leq \alpha.
\end{array}\right.
\end{IEEEeqnarray}
\end{cor}
This formula is the same as the one given in \cite{PBT} with $M$ and $N$ interchanged, which makes sense considering the reciprocity result in Corollary~\ref{cor_reciprocity_GDOF}. In other words, $d_{sym}$ of the $(M,N,M,N)$ IC with $M>N$ for a given $\frac{M}{N}=r$ is the same as that of the GDoF of a MIMO IC with $M\leq N$ and $\frac{M}{N}=\frac{1}{r}$. However, the achievable schemes on the two channels are entirely different. While for $M\leq N$ the coding scheme need not depend on the channel matrices at the transmitters (see the achievability scheme of \cite{PBT}), for $M>N$ the covariance matrices are necessarily functions of the channel matrices. Hence, a naive extension of the scheme of \cite{PBT} to the case of $M>N$ is not GDoF optimal (it wouldn't even be DoF optimal). Figure~\ref{figure_gdof_MgeqN_channel} and \ref{figure_gdof_PBT_scheme} shows the GDoF achievable by the coding scheme of this paper and the coding scheme of~\cite{PBT} for $M>N$ (which we refer to as the PBT scheme). The dashed red line indicates the GDoF of the scheme that treats interference as noise (referred to as TIN).

\section{Some insights}
\label{sec_insights}
\vspace{-1.25mm}
\subsection{Only Tx/Rx ZF Beam-forming is not GDoF optimal}
\label{subsec_BFZF_suboptimality}
\begin{figure}[htp]
  \begin{center}
  \subfigure[$\alpha=\frac{2}{3}$.]{\label{figure_gdof_3232-a}\includegraphics[scale=0.18]{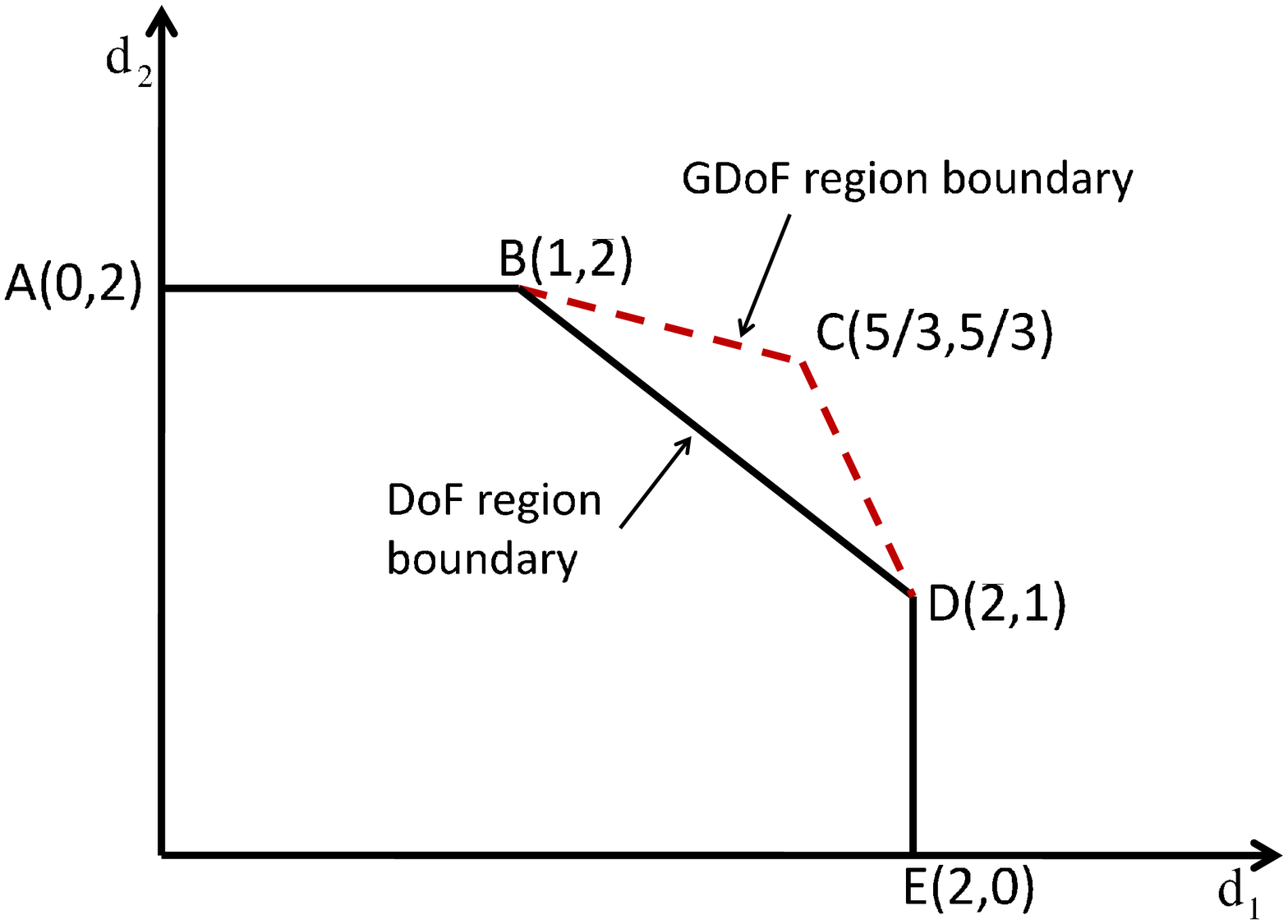}}
  \subfigure[$(1,1,2,1)$ IC.]{\label{figure_ZF_channel-b}\includegraphics[scale=.15]{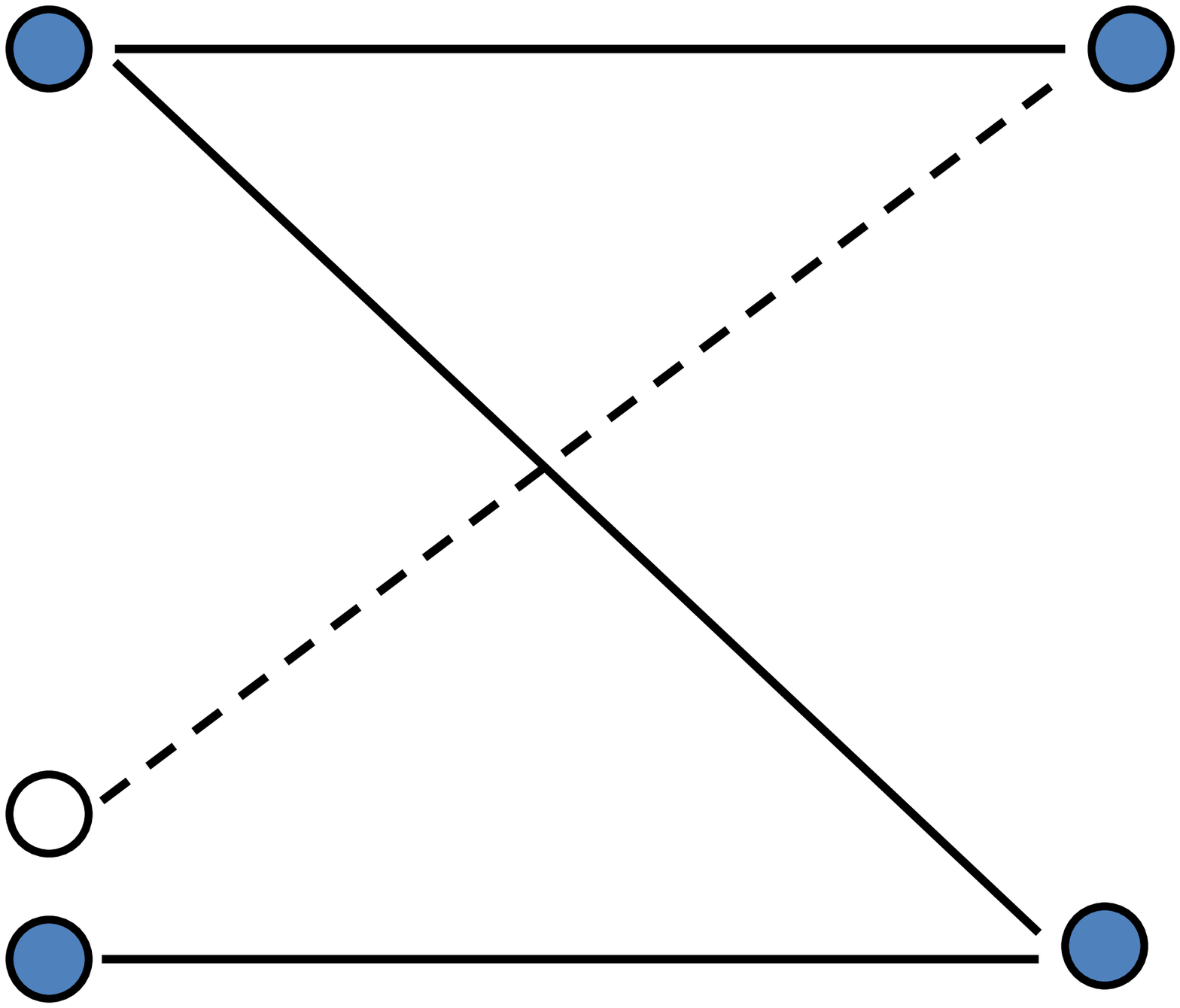}}
  \end{center}
\caption{Diagonalization of the cross links using ZF to achieve the DoF region. }
\label{figure_gdof_3232}
\end{figure}

The fundamental GDoF is a finer high SNR approximation and therefore reveals insights that are not revealed by the DoF analysis. Figure~\ref{figure_gdof_3232-a} illustrates this point by comparing the DoF and GDoF region of the $(3,2,3,2)$ IC with $\bar{\alpha}=[1,\frac{2}{3},\frac{2}{3},1]$. It is well known from \cite{JFak} that only transmit/receive zero-forcing beam-forming is sufficient to achieve any point in the DoF region of the channel. This GDoF achievable using this scheme is shown in Fig. \ref{figure_gdof_3232} as against the fundamental GDoF. It is easily seen that forgoing the opportunity to align signals in the signal-level dimension leads to a strictly GDoF suboptimal performance. In particular, this technique can not achieve any point in the triangular region BCD. However, the coding scheme of Section~\ref{sec_channel_model_and_preliminaries} which in addition to beamforming utilizes signal-level interference alignment can achieve all the points in the region BCD.


\vspace{-1.25mm}
\subsection{On achieving single user performance}
\label{subsection_single_user_perf}
It is well known that the achievability of single user DoFs on a MIMO MAC or BC depends on the number of antennas at the different nodes and on a SISO IC it depends on the interference level, $\alpha$. On a MIMO IC it depends on both. From Corollary~\ref{cor_gdof_symmetric_MgeqN} we get that on a $(M,N,M,N)$ IC with $M\geq N$, the single user GDoF is achieved (each user can gets $N$ DoFs) when
\begin{equation*}
    \alpha\geq \alpha^*=\left(3-\frac{M}{N}\right).
\end{equation*}
In contrast to the case on a SISO IC, the value of $\alpha$, at which the single user performance is achieved, denoted by $\alpha^*$, decreases below $2$ as $M$ increases, giving a similar effect as in a MAC or BC (see Fig. \ref{figure_gdof_MgeqN_channel}).

\vspace{-1mm}
\subsection{Sub-optimality of treating interference as noise}
Another fundamental difference of the MIMO IC from the SISO IC revealed by the GDoF analysis is this: in general, treating interference as noise (TIN) is {\em not} GDoF optimal on a MIMO IC even in the {\it very weak interference} regime, i.e., when $\alpha\leq \frac{1}{2}$. This is seen in Fig. \ref{figure_gdof_MgeqN_channel} where the dotted line, which represents the symmetric GDoF achievable by TIN, is strictly sub-optimal with respect to the fundamental GDoF of the channel for $\alpha\leq \frac{1}{2}$ whenever $M/N>1$. See also Fig. \ref{figure_gdof_3223_channel}.

\vspace{-1.25mm}
\subsection{Deviation from the ``W" shape}
\begin{figure}[htp]
  \begin{center}
    \subfigure[Symmetric GDoF region of the $(3,2,2,3)$ IC.]{\label{figure_gdof_3223_channel}\includegraphics[scale=.28]{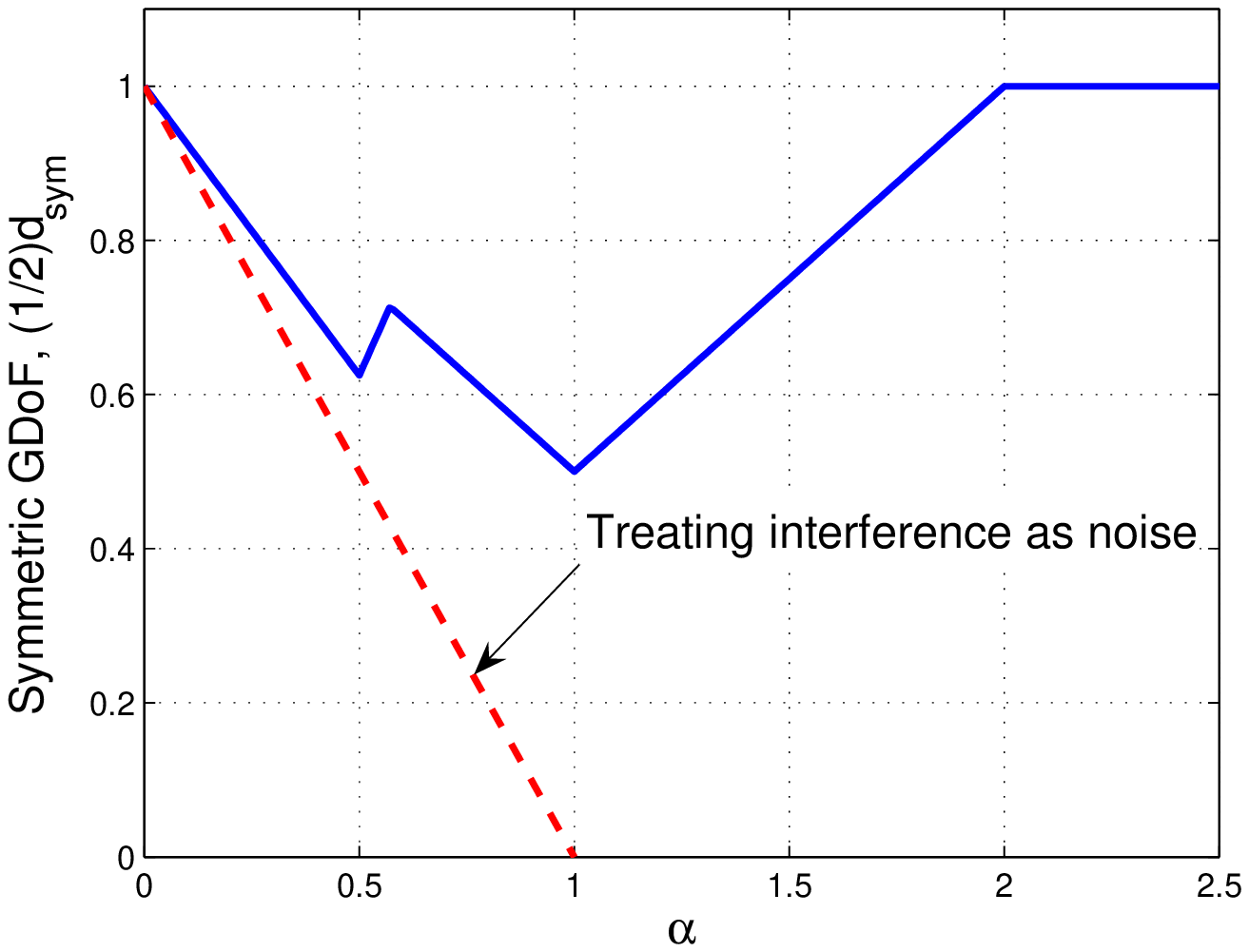}}
    \subfigure[Symmetric GDoF region of a $(1,1,2,1)$ MIMO IC.]{\label{figure_gdof_1121}\includegraphics[scale=0.28]{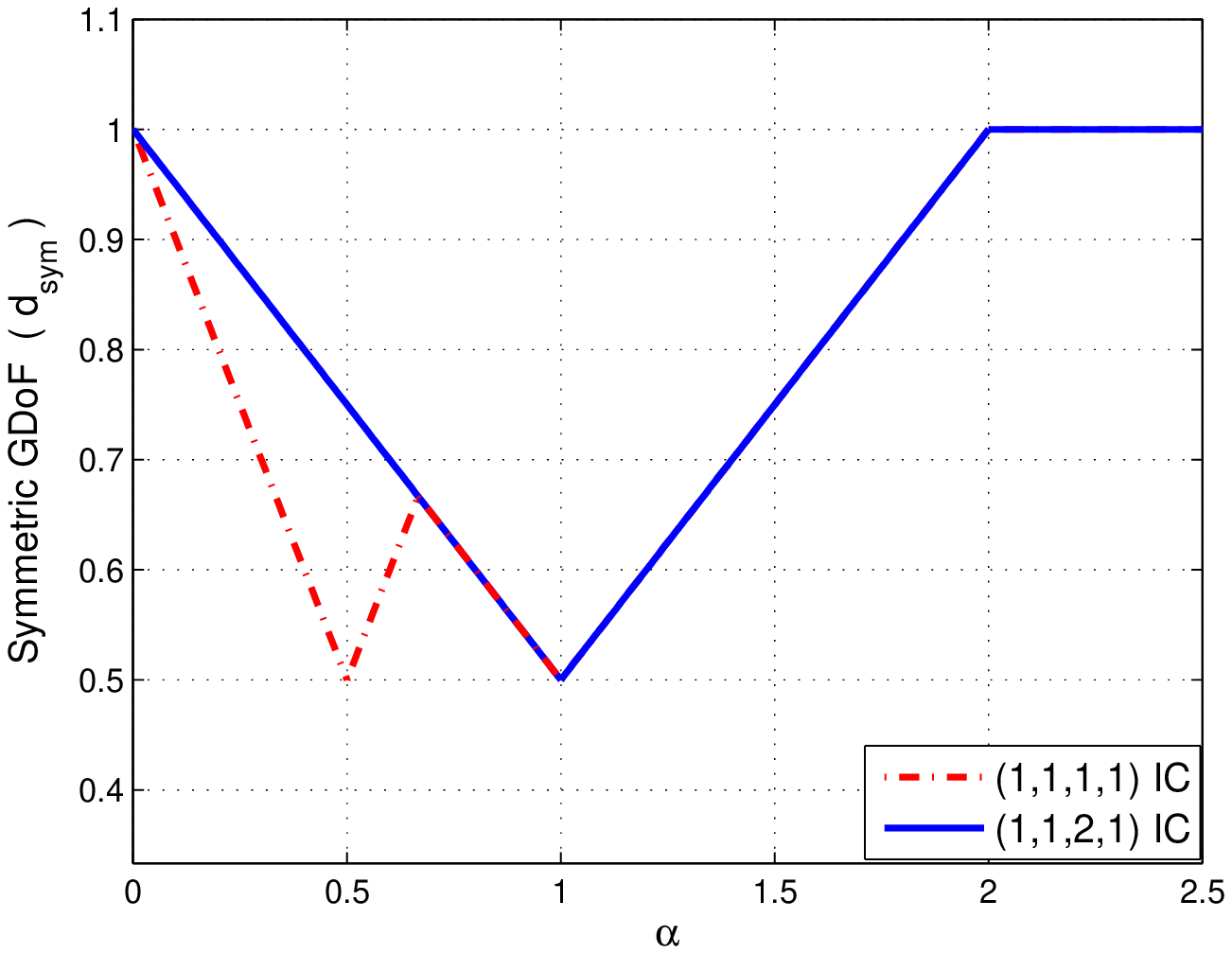}}
  \end{center}
\caption{Sub-optimality of TIN and deviation of the GDoF boundary from the well known ``W" shape. }
\end{figure}

Unlike in the SISO IC, the symmetric GDoF region of a MIMO IC in general need not maintain the ``W" shape. The deviation in general is due to asymmetry in the numbers of antennas. For example, consider the $(1,1,2,1)$ IC with $\alpha_{ii}=1$ and $\alpha_{ij}=\alpha$, for $i\neq j\in \{1,2\}$. The best achievable symmetric DoF ($d_{sym}=d_1=d_2$) on this channel denoted by $d$ is
\begin{equation*}
d=\left\{\begin{array}{ll}
1-\frac{\alpha}{2}, &0\leq \alpha\leq 1;\\
\frac{\alpha}{2}, & 1\leq\alpha\leq 2;\\
1, & 2\leq \alpha.
\end{array}\right.
\end{equation*}
which is depicted in Figure~\ref{figure_gdof_1121}. Diagonalizing the cross-link from $Tx_2$ to $Rx_1$ and then turning off the subchannel which interferes with $Rx_1$ gives the GDoF equivalent channel of Figure~\ref{figure_ZF_channel-b} which is a SISO ``Z" IC. The symmetric GDoF region of this channel is indeed ``V" shaped as found in \cite{ETW1}. Although a little more involved, the distorted ``W" of Figure~\ref{figure_gdof_3223_channel} can be explained similarly.

\section{Conclusion}
\label{sec_conclusion}

The GDoF analysis of this paper, unifies the earlier results on GDoF of SISO IC~\cite{ETW1}, the DoF region~\cite{JFak} of MIMO IC and the symmetric GDoF~\cite{PBT} of MIMO IC through a single achievable scheme for all. The coding schemes in~\cite{JFak} and \cite{PBT} are strictly suboptimal in the GDoF sense on a general 2-user MIMO IC in one case or other. The analysis here reveals various insights about the MIMO IC including the fact that in general, partially decoding the unintended user's message is necessary to be GDoF optimal even in the so called very weak interference regime. The two types of signaling dimensions available on a MIMO IC -- namely, signal space and signal level -- are jointly exploited in the GDoF optimal scheme.

\bibliographystyle{IEEEtran}
\bibliography{mybibliography}
\end{document}